\documentclass[12pt]{article}
\usepackage{epsf}
\usepackage{color}
\usepackage{graphicx}
\begin{document}
\newcommand{\ccdot}{ \! \cdot \! }
\newcommand{\Br}{{\cal B}}
\newcommand{\beq}{\begin{equation}}
\newcommand{\eeq}{\end{equation}}
\newcommand{\la}{\langle}
\newcommand{\cO}{{\cal O}}
\newcommand{\ra}{\rangle}
\newcommand{\beqa}{\begin{eqnarray}}
\newcommand{\eeqa}{\end{eqnarray}}
\newcommand{\no}{\nonumber}
\newcommand{\yuk}{{Y}}
\def\Green{} 
\definecolor{mygreen}{rgb}{0,0.6,0}
\newcommand{\green}[1]{\textcolor{mygreen}{#1}}
\newcommand{\blue}[1]{\textcolor{blue}{#1}}
\newcommand{\red}[1]{\textcolor{red}{#1}}
\def\Black{}

\def\ap#1#2#3{     {\it Ann. Phys.  }{\bf #1} (#2) #3}

\def\arnps#1#2#3{  {\it Annu. Rev. Nucl. Part. Sci. }{\bf #1} (#2) #3}
\def\npb#1#2#3{    {\it Nucl. Phys. }{\bf B #1} (#2) #3}
\def\plb#1#2#3{    {\it Phys. Lett. }{\bf B #1} (#2) #3}
\def\pr#1#2#3{     {\it Phys. Rev. }{\bf   #1} (#2) #3}
\def\prd#1#2#3{    {\it Phys. Rev. }{\bf D #1} (#2) #3}
\def\prb#1#2#3{    {\it Phys. Rev. }{\bf B #1} (#2) #3}
\def\prep#1#2#3{   {\it Phys. Rep. }{\bf #1} (#2) #3}
\def\prl#1#2#3{    {\it Phys. Rev. Lett. }{\bf #1} (#2) #3}
\def\ptp#1#2#3{    {\it Prog. Theor. Phys. }{\bf #1} (#2) #3}
\def\rpp#1#2#3{    {\it Rept. Prog. Phys. }{\bf #1} (#2) #3}
\def\ppnp#1#2#3{   {\it Prog. Part. Nucl. Phys. }{\bf #1} (#2) #3}
\def\rmp#1#2#3{    {\it Rev. Mod. Phys. }{\bf #1} (#2) #3}
\def\zpc#1#2#3{    {\it Z. Phys. }{\bf C #1} (#2) #3}
\def\ijmpa#1#2#3{  {\it Int. J. Mod. Phys. }{\bf A #1} (#2) #3}
\def\sjnp#1#2#3{   {\it Sov. J. Nucl. Phys. }{\bf #1} (#2) #3}
\def\yf#1#2#3{     {\it Yad. Fiz. }{\bf #1} (#2) #3}
\def\jetpl#1#2#3{  {\it JETP Lett. }{\bf #1} (#2) #3}
\def\epjc#1#2#3{   {\it Eur. Phys. J. }{\bf C #1} (#2) #3}
\def\ibid#1#2#3{   {\it ibid. }{\bf #1} (#2) #3}
\def\jhep#1#2#3{   {\it JHEP  }{\bf #1} (#2) #3} 
\def\physica#1#2#3{{\it Physica }{\bf A #1} (#2) #3}
\def\ncim#1#2#3{   {\it Nuovo Cim. }{\bf #1} (#2) #3}

\hfill  INFNNA-IV-2003/15
\renewcommand{\thefootnote}{\fnsymbol{footnote}}
\begin{center}
{\LARGE \bf  Recent developments in rare kaon decays}\footnote[1]{To appear in
Modern Physics Letters A (MPLA)}
\end{center}
\vspace{0.8cm}
\centerline{ Giancarlo D'Ambrosio}

\centerline{\it INFN-Sezione di Napoli, I-80126 Napoli, Italy}
\centerline{\emph{E-mail: giancarlo.dambrosio@na.infn.it}}


\baselineskip=11.6pt
\begin{abstract}
We discuss  theoretical issues in rare and  radiative  kaon decays.
The interest is twofold: to extract useful short-distance information
and understand the underlying dynamics. 
We emphasize channels where either we can understand non-perturbative 
aspects of QCD or  there is a chance to test the Standard Model.
\end{abstract}
\baselineskip=14pt
\section{Introduction}
The  Standard Model (SM) is in very good shape. However some questions, 
like a satisfactory 
solution of the hierarchy problem, do not have yet a satisfactory answer. 
Also some very decisive tests of the SM, like the $g-2$ or  the  amount of 
direct CP violation in $K_L \rightarrow \pi \pi$ are plagued by uncertainties
related to our ignorance on hadronic matrix elements.
Thus our goal is to show that in rare kaon decays 
\cite{reviews,Donoghuebook,Littenberg:2002um,DI98}
there are: i) {\it Golden modes}, like 
$K\rightarrow \pi \nu \overline{\nu }$ \cite{buchalla96}, completely dominated
by short distance, where 
the SM is challenged to a very meaningful level, ii) channels,
like $K_L \rightarrow \pi ^0  e \overline{e}$ and 
$K_L \rightarrow   \mu  \overline{\mu}$ , where our good knowledge of long 
distance dynamics allows us to single out quite accurately
the  interesting short distance dependence \cite{rad01}, iii) channels like 
$K_S \rightarrow \gamma \gamma $, completely dominated by long distance
\cite{DEG}  but 
accurately predicted by chiral perturbation theory ($\chi$PT), and thus they are both
 important tests of the theory and also, as we shall see,  relevant
complementary channels to 
$K_L \rightarrow \pi ^0  e \overline{e}$.

$B$-physics will test the SM measuring the CKM triangle \cite{reviews} 
with sizes $%
V_{qb}^{*}V_{qd};$ the area of this triangle, $J_{CP}/2,$ is invariant for
all CKM\ triangles and non-zero if $CP$ is violated; in the Wolfenstein
parametrization:\ 
\begin{equation}
\left| J_{CP}\right| \stackrel{Wolfenstein}{\simeq }A^{2}\lambda ^{6}\eta
\label{eq:j}
\end{equation}
with $V_{us}=\lambda ,$ $V_{cb}=$ $A\lambda ^{2},\Im m(V_{td})=-A\lambda
^{3}\eta $. 
As we shall see, as a consequence of our improved understanding of low energy
physics we can test precisely (\ref{eq:j}) in rare kaon channels
\cite{reviews}.
This is particularly exciting since there are several experiments 
aiming the required
accuracy
\cite{Littenberg:2002um}. 
Since we will use some $\chi$PT results and to illustrate the
relevance of the recent NA48 measurement of $K_S \rightarrow \gamma \gamma $
\cite{Lai:2002sr}, we will briefly mention some $\chi$PT achievements
and then discuss respectively 
$ K\rightarrow \pi \nu \overline{\nu }$,  $K\rightarrow \pi e\overline{e}$ and
 the related decays $K \rightarrow \pi \gamma \gamma $.

\section{Chiral Perturbation Theory and $K_S \rightarrow \gamma \gamma $}
QCD is non-perturbative at energy scales below $1$ GeV and thus symmetry  
arguments must be invoked in order to be predictive. 
QCD for massless quarks  exhibits the global symmetry $%
SU(3)_{L}\otimes SU(3)_{R}$ and there are strong phenomenological 
arguments (Gold\-ber\-ger-Trei\-man relation, ...) that the pion is 
the Goldstone boson
of the broken symmetry  
$SU(3)_{L}\otimes SU(3)_{R}\rightarrow SU(3)_{V}$. 
Thus  $\chi PT$ \cite{Donoghuebook,DI98,Weinberg1,chiralrev} is an
effective field theory based on the following two assumptions: i) the
pseudoscalar mesons are the Goldstone bosons (G.B.) of the symmetry above, 
ii) there
is a {\it (chiral) power counting}, i.e. the theory has a small expansion
parameter: $p^{2}/$ $\Lambda _{\chi SB}^{2}$ and/or $m^{2}/$ $\Lambda _{\chi
SB}^{2},$ where $p$ is the external momenta, $m$ the masses of the G.B.'s
and $\Lambda _{\chi SB}$ is the chiral symmetry breaking scale: $\Lambda
_{\chi SB}\sim 4\pi F_{\pi }\sim 1.2$ GeV. Being an effective field theory,
loops and counterterms are required by unitarity and have to be evaluated
order by order
\cite{Donoghuebook,Weinberg1,chiralrev}. 
It turns out more practical to descibe the 
chiral fields through a non-linear realization, $U=e ^{i\sqrt{2} \Phi/F}$ and 
$ \Phi=\sum_i \lambda _i \phi ^i $, $\lambda _i$ are the Gell-Mann matrices,
 $F\sim F_\pi$.
We can split the lagrangian in a strong 
($\Delta S=0$) and in a weak non-leptonic piece ($\Delta S=1$):
 ${\cal L} =  {\cal L} _{\Delta S=0} + {\cal L} _{\Delta S=1}$, and 
then consider the chiral expansion 
\beq
{ {\cal L}} _{\Delta S=0} ={ {\cal L}} _{\Delta S=0} ^2
+{ {\cal L}} _{\Delta S=0} ^4  +\cdots 
 ={F^2 \over 4} \underbrace{
\la D_\mu U D^\mu U^\dag + \chi U^\dag + U\chi ^\dag \ra}_{\pi\rightarrow l \nu,\ 
\pi \pi \rightarrow \pi \pi} +\sum_i  L_i O_i +\cdots \label{eq:LS}
\eeq
\beq
{ {\cal L}} _{\Delta S=1}={ {\cal L}} _{\Delta S=1} ^2
+{ {\cal L}} _{\Delta S=1} ^4  +\cdots  = {G_8 F^4 }
 \underbrace{
\la \lambda _6 
D_\mu U^\dag D^\mu U \ra}_{K \rightarrow 2 \pi/3\pi } +
\underbrace{{G_8 F^2 } \sum_i N_i W_i}_{K^+ \rightarrow \pi^+
 \gamma \gamma, \ K \rightarrow \pi \pi \gamma} +\cdots \label{eq:LW}, 
\eeq
where $\chi$ is the appropriate $SU(3)$-spurion that generate the G.B. masses
and 
the second  terms in  (\ref{eq:LS}) and  (\ref{eq:LW}) represent
respectively the  strong  ${\cal O} (p^4)$ (Gasser-Leutwyler
\cite{Weinberg1}) and the weak   ${\cal O} (p^4)$ \cite{EKW93,DP98} 
lagrangian. One of the most fantastic $\chi PT$ predictions is the
 $\pi \pi$-scattering lenght \cite{DI98,Weinberg1,chiralrev}
in terms only of the pion decay constant, $ F_\pi$: the ${\cal O} (p^2)$
result in Eq. (\ref{eq:LS}), describing simultaneously 
the pion kinetic term and pion interactions,
 is phenomenologically correct up to $30\%$ corrections, fully
predicted by the higher orders \cite{GC00}.

\begin{figure}[t]
\begin{center}
\leavevmode
\epsfysize=4 cm
\epsfxsize=15 cm
\epsfbox{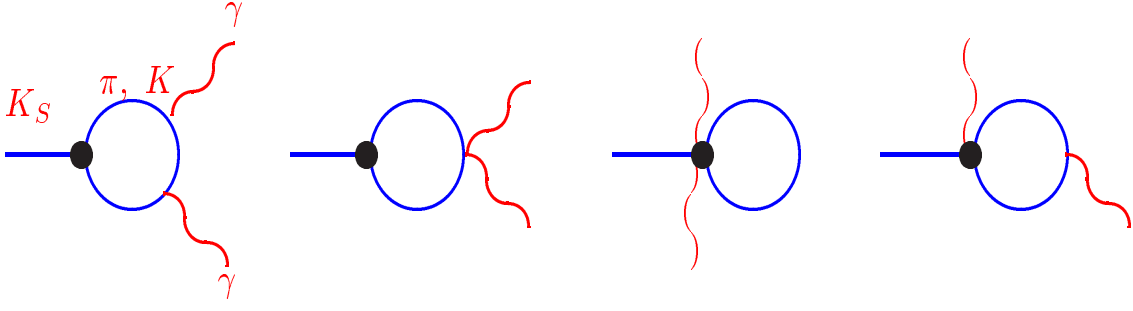}
\end{center}
\caption{$K_S\rightarrow \gamma \gamma $: the $\bullet$  represents
the  ${\cal O }(p^2)$ weak vertex, proportional to
 $G_8$, defined in (\ref{eq:LW}), \cite{DEG}}
\label{fig:KSgg} 
\end{figure}

$K_S\rightarrow\gamma\gamma$ has vanishing short-distance 
contributions and thus it is a pure 
long-distance phenomenon; since the external particles are neutral there is no 
${\cal O }(p^2)$ amplitude. For the same reason, if we write down the
 ${\cal O}(p^4)$ counterterm structure,
$F_{\mu\nu}F^{\mu\nu} \langle\lambda_6 Q U^+ Q
U\rangle$, this gives a vanishing contribution.
This implies that at ${\cal O}(p^4)$:  i) we 
have only  a loop contribution in Fig. \ref{fig:KSgg}
 and ii) this contribution is scale-independent \cite{DEG}:
\beqa
A(K_S\rightarrow {\gamma}  {\gamma}) =& \displaystyle{ {2\alpha F
\over \pi M^{2}_{K} }}
( G_8 +{2\over 3}G_{27} )\, ( {M}^{2}_{K} - {M}^{2}_{\pi})\cdot&\hskip-0.3cm  
\left(1+\frac{{M}^{2}_{K}}{{M}^{2}_{\pi}}
\ln ^2 \frac{\beta -1}{1+\beta}\right)\cdot\hskip2cm\no \\
&\big[ (q_{1}{\epsilon}_{2})(q_{2} {\epsilon}_{1})
-({\epsilon}_{1}{\epsilon}_{2})({q}_{1} {q}_{2})\big], &\hskip1cm 
\beta= \sqrt{1-4{M}^{2}_{\pi}/{M}^{2}_{K}}\hskip2cm
\label{eq:ksgga}
\eeqa
where $G_8$, defined in (\ref{eq:LW}) and $G_{27}$, the coefficient of 
$\Delta I=3/2$-transitions are  completely predicted
by the  $K\rightarrow \pi \pi $ amplitudes. This is the {\it ideal} test
of  $\chi PT$ 
(and in general of effective field theories) at the {\it quantum level}.
At  higher order,
${\cal O }(p^6) $, $\pi$-loop corrections are  small \cite{KH94},  while
contributions to $A^{(6)}$
 from  ${\cal L} _{\Delta S=1} ^6$,  are chirally suppressed: 
\begin{equation}
{\cal L} _{\Delta S=1} ^6\supset\displaystyle{\frac{c}{(4\pi F_\pi)^2}}
F^{\mu\nu}F_{\mu\nu}\langle\lambda _6 Q^2 \chi
U^+\rangle \qquad \Rightarrow \displaystyle{\frac{A^{(6)}}{A^{(4)}}} \sim
\displaystyle{\frac{m_K^2}{(4\pi F_\pi)^2}}\sim 
 0.2\label{eq:Ksp6},
\end{equation}
where  $c \sim {\cal O}(1)$ has to be determined phenomenologically but 
has no vector Meson (VMD) contributions and so it is not enhanced by the
factor $(4\pi F_\pi)^2 /m_V ^2 \sim (1200/770)^2$. $Q$ is the diagonal quark
electric charge matrix: $Q=diag(2/3,-1/3,-1/3)$.
So  we can compare 
the ${\cal O }(p^4) $ prediction in  (\ref{eq:ksgga})
 with the recent
NA48 result \cite{Lai:2002sr}:
\begin{equation}
\Br (K_S\rightarrow\gamma\gamma)=\left\{\begin{array}{l}
{\rm TH} \quad  (p^4) \quad  2.1\times 10^{-6}\\
\\
{\rm NA48} \quad (2.78\pm 0.072)\times 10^{-6}\\ \end{array} 
\quad \Rightarrow \displaystyle{\frac{A^{(6)}}{A^{(4)}}} \sim 15\%
\right.
\end{equation}
The error in the amplitude,  is a success of 
the na\"{i}ve expectation
in (\ref{eq:Ksp6}), and fixes $c$ in  Eq.(\ref{eq:Ksp6}).

\section{$K\rightarrow \pi \nu \overline{\nu }$}

The SM predicts the $V-A\otimes V-A$ effective hamiltonian 
\beqa
{\cal H}& = &\frac{G_{F}}{\sqrt{2}} \frac{\alpha}{2 \pi \sin ^2 \theta_W}
(\ \underbrace{V_{cs}^{*}V_{cd}\ X_{NL}}_{\textstyle{\lambda x_c}} \,
+ \underbrace{ V_{ts}^{*}V_{td}X(x_t)}_{\textstyle{A^2\lambda ^5 \ 
(1-\rho -i{\eta}){ x_t }}}) \ \overline{
s}_L \gamma _\mu d_L \  \overline{\nu } _L \gamma ^\mu \nu _L , 
\label{ampsd} \\ \no
\eeqa
 $x_q=m_q^2/M_W^2$, $\theta_W$ the Weinberg angle and 
 $X$'s are the Inami-Lin functions with 
Wilson coefficients known at next--to--leading order \cite{buchalla96}. 
$SU(2)$ isospin symmetry relates hadronic matrix elements for $%
K\rightarrow \pi \nu \overline{\nu }$ to $K\rightarrow \pi l\overline{\nu }$
to a very good precision \cite{Marciano96} while long distance contributions are negligible
\cite{gino98a}. QCD corrections have been
evaluated at next-to-leading order \cite{reviews,buchalla99b} and 
the main uncertainties
is due to the strong corrections to the charm loop contribution.
\begin{figure}[t]
\begin{picture}(640,140)(0,0)
\put(55,155){{\green{${\scriptstyle 
\Br (K^{^{+ }}\rightarrow \pi ^{+ }\nu 
\overline{\nu })}$: 1 ${\scriptstyle \sigma}$ 
{\footnotesize lower limit}}}}
\thicklines
\put(60,150){\green 
}\multiput(70,150)(3.5,0){27}{\green {\circle*{1}}}
\put(170,150){{\green {\vector(-1,0){10}}}}
\thinlines
\put(10,25){$\epsilon$}
\put(10,115){{\blue{$\sin 2\beta$}}}
\put(170,50){\parbox{23mm}{\scriptsize \red{$68\%$} and \red{$90\%$} C.L.
{\red{ellipses}} from the global fit  }}
\put(205,138){\parbox{22mm}{{\blue{ 
${\scriptscriptstyle \Delta M_{B_d}/\Delta M_{B_s}}$
{\scriptsize $90\%$  limit} }}}}
\thicklines
\put(0,0){\includegraphics[width=7.5cm,height=6cm]{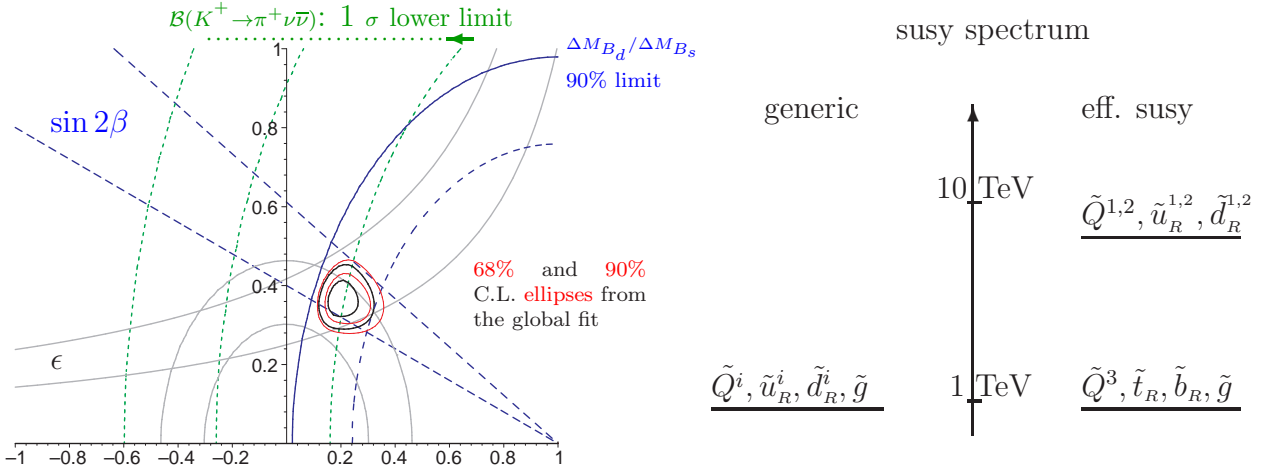}}
\put(355,0){ {\vector(0,1){125}}}
\put(330,150){susy spectrum}
\put(280,120){generic}
\put(400,120){eff. susy}
\put(350,15){1 TeV }
\put(357,13){\line(1,0){5}}
\put(357,88){\line(1,0){5}}
\put(345,90){10 TeV}
\put(260,10){\line(1,0){65}}
\put(260,15){$\tilde{Q^i},\tilde{u}_{\scriptscriptstyle R}^i,
\tilde{d}_{\scriptscriptstyle R}^{i},\tilde{g}$}
\put(400,10){\line(1,0){60}}
\put(400,15){$\tilde{Q}^3,\tilde{t}_{\scriptscriptstyle R},\tilde{b}_{\scriptscriptstyle R},\tilde{g}$}
\put(400,75){\line(1,0){60}}
\put(400,80){$\tilde{Q}^{1,2},\tilde{u}_{\scriptscriptstyle R}^{^{1,2}},
\tilde{d}_{\scriptscriptstyle R}^{^{1,2}}$}
\end{picture}
\caption{On the left side we show  in the two ellipses the
allowed region in the  $\bar\rho-\bar \eta$ plane by the global fit
without 
imposing the $\Br(K^+\to\pi^+ \nu \bar\nu)$; the dotted curves on the left define the 
$\Br (K^{^{+ }}\rightarrow \pi ^{+ }\nu 
\overline{\nu })$: 1 ${ \sigma}$ lower bound \cite{D'Ambrosio:2001zh}: 
the mismatch
between  the central value 
implied by $\Br (K^{^{+ }}\rightarrow \pi ^{+ }\nu 
\overline{\nu })$ and the ellipses of the global fit
 can be also  interpreted in terms of NP in  $\Delta B=2$ transitions,
 then the effective
supersymmetry scenario  \cite{Dimopoulos:1995mi} (right side) may be appealing
\cite{D'Ambrosio:2001zh}, $\tilde{Q}^i$ are the $SU(2)_L$ quark doublets}
\label{fig:2}
\end{figure}
The structure in (\ref{ampsd}) leads to a pure CP violating contribution
to $K_{L}\rightarrow \pi ^{0}\nu \overline{\nu },$ induced only from the top
loop contribution and thus proportional to $\Im m(\lambda _{t} )$
($\lambda _t= V_{ts}^{*}V_{td}$) and free of
hadronic uncertainties. This leads to the prediction \cite{buchalla96}
\begin{equation}
\Br (K_{L}\rightarrow \pi ^{0}\nu \overline{\nu })_{SM}=4.25\times
10^{-10}\left[ \frac{\bar{m}_{t}(m_{t})}{170GeV}\right] ^{2.3}\left[ \frac{%
\Im m(\lambda _{t})}{\lambda ^{5}}\right] ^{2}.  \label{eq:klpi0nunu}
\end{equation}
$K^{^{\pm }}\rightarrow \pi ^{\pm }\nu \overline{\nu }$ receives CP
conserving and violating contributions proportional to $\Re e(\lambda _{c}),$
$\Re e(\lambda _{t})$ and $~~\Im m(\lambda _{t}).$ Theoretical uncertainty
from the charm loop induces 8\% error on the width. If one takes into
account the various indirect limits, i.e.$V_{ub}$ and\ $\varepsilon ,$ on
CKM elements one obtains the SM values \cite{reviews,buchalla96,CKMfits}: 
\begin{equation}
\Br (K_{L}\rightarrow \pi ^{0}\nu \overline{\nu })\ =\left( 2.8\pm 1.0\right)
\times 10^{-11}\quad \Br (K^{\pm }\rightarrow \pi ^{\pm }\nu \overline{\nu }%
)=\left( 0.72\pm 0.21\right) \times 10^{-10} \label{eq:kpnnSM}
\end{equation}
Two events have been observed by E787 \cite{E787} leading to  
\begin{equation}
\Br(K^{^{\pm }}\rightarrow \pi ^{\pm }\nu \overline{\nu })=\left(
1.57_{-0.82}^{+1.75}\right) \times 10^{-10} \qquad {\rm E787}\label{eq:E787}
\end{equation}
The direct existing upper bound  for the neutral decay:
$\Br (K_{L}\rightarrow \pi ^{0}\nu \overline{\nu })\ \leq 5.9\times 10^{-7}$
\cite{kpi0pi0nunu}, can  be improved according the following consideration:
 the isospin structure of any $\overline{s}d$ operator (bilinear in the quark
fields) leads to the model independent relation among 
$A(K_{L}\rightarrow \pi ^{0}\nu \overline{\nu } )$ and 
$A(K^{^{\pm }}\rightarrow \pi ^{\pm }\nu \overline{\nu })$
\cite{GNir97} and to the
interesting bound with E787 result \cite{E787} 
\[
\Br(K_{L}\rightarrow \pi ^{0}\nu \overline{\nu })<\frac{\tau _{K_{L}}}{\tau
_{K^{+}}}\ \Br (K^{^{\pm }}\rightarrow \pi ^{\pm }\nu \overline{\nu }) 
\begin{array}{c}
< \\ 
E787
\end{array}
1.7\cdot 10^{-9}\quad {\rm at\quad }90\%C.L. 
\] 
Future measurements: i) $K^+$, BNL (E787 and E949) should improve the present 
result, while  CKM at Fermilab should measure the branching with a 
 10\% accuracy, ii) $K_L$, KOPIO at BNL and KEK are aiming to measure this
channel \cite{Littenberg:2002um}.

\begin{figure}
\begin{picture}(640,140)(0,0)
\put(50,0){\includegraphics[width=4cm,height=4cm]{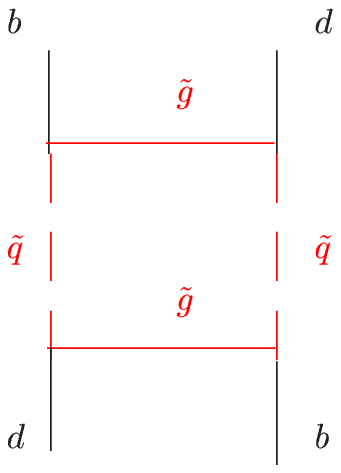}}
\put(270,0){\includegraphics[width=5.5cm,height=4cm]{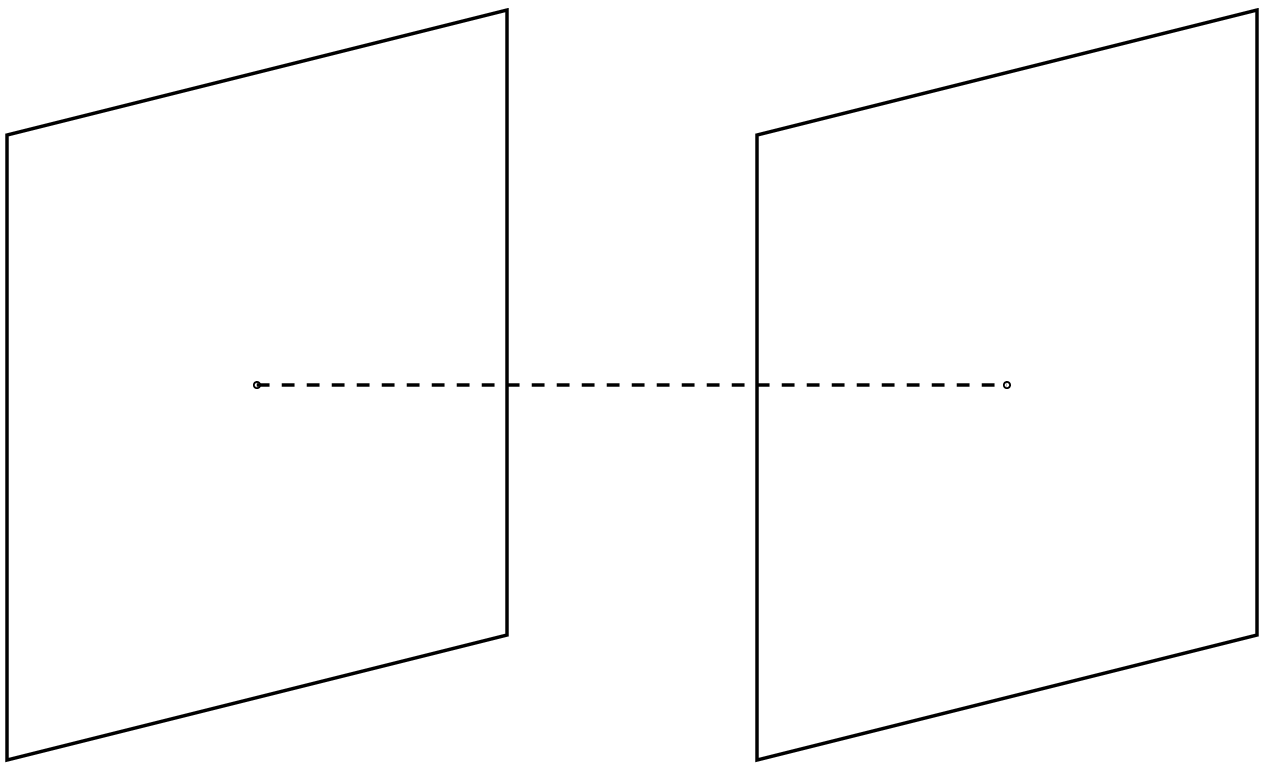}}
\put(280,85){$G_F$}
\put(375,85){$SM$-fields}
\put(340,65){$\phi$}
\end{picture}\label{fig:3}
\caption{On the left side,${\cal H}_{\Delta F=2}$ generated by  gluino exchange. On the right side
MFV in the contest of extra dimensions}
\end{figure}

One can speculate that the central value in (\ref{eq:E787}) is overshooting
the SM prediction in (\ref{eq:kpnnSM}) \cite{D'Ambrosio:2001zh} and NP 
is required. Referring to the original reference (and \cite{Isidori:2003qc})
for a detailed
discussion, we  show in Fig. \ref{fig:2} 
the $K^{^{\pm }}\rightarrow \pi ^{\pm }\nu \overline{\nu }$
preferred values $\bar{\rho}-\bar{\eta}$ versus the values 
allowed by $\sin(2\beta)$, $\epsilon$ and $\Delta M_{B_d}/\Delta M_{B_s}$.
 Then two possibilities are envisaged in order to reconciliate the 
central value
$\Br (K^{^{\pm }}\rightarrow \pi ^{\pm }\nu \overline{\nu })$
(see Fig. \ref{fig:2})
with the SM prediction in (\ref{eq:kpnnSM}):
i) NP in $\overline{s}\rightarrow \overline{ d}\overline{\nu}\nu$   
implying NP in $\overline{b}\rightarrow \overline{ s}\overline{\nu}\nu$
or ii) NP in $\Delta B=2$. This last possibility seems also motivated
by other $B$-observables, like $B\to\pi K$ \cite{Fleischer:2003xx}.
However it is not harmless to add FCNC's. Let's look the SM Yukawa structure
\beq
{\cal L}^\yuk _{SM}\  ~ = ~ \  {\bar Q} \yuk_D D  H
+ {\bar Q} {\yuk_U} U  H_c
+ {\bar L} {\yuk_E} E  H {\rm ~+~h.c.}~\label{eq:yuk}
\eeq
where $Q,U,D$  ($L,E$) are respectively the quark (lepton) doublets and singlets.
Diagonalization of the quark matrices lead to the CKM unitary matrix, $V_{ij}$ and
GIM mechanism for FCNC's. 
Neglecting strong corrections
\beq
{\cal H} _{\Delta F=2}^{SM}\sim \frac {G_F^2 M _{W}^2}{16\pi^2} 
\left[\frac{(V_{td}^* m _t ^2 V_{tb})^2}{v^4}
(\bar{d}_L \gamma ^\mu b_L)^2 + \frac{(V_{td}^* m _t ^2 V_{ts})^2}{v^4}
(\bar{d}_L \gamma ^\mu s_L)^2\right]+{\rm charm,up}\label{eq:HSM}\eeq
In supersymmetry new flavour structures are generated by the soft mass terms: 
\beq {\cal L} _{soft}=
\tilde{Q} ^\dag m_Q ^2 \tilde{Q}+ \tilde{L}^\dag  m_L^2 \tilde{L}+
\tilde{\bar{U}}a_u \tilde{Q}H_u+\cdots \label{eq:soft}\eeq
The diagonalization of these contributions add new flavour matrices to the CKM $V$ in
 (\ref{eq:HSM}). For instance assuming the dominance of
 the $LL$  gluino-sdown box diagrams in Fig. \ref{fig:3} \cite{GGMS96}
 \beq
{\cal H}^{ \tilde{g}}_{\Delta F=2}  \sim   \frac{\alpha _s ^2}{9 M^2_{\tilde{Q}} }
\left[(\delta ^{LL} _{12})^2 ~  (\bar{s}_L\gamma_\mu d_L)^2+\cdots
\right]\qquad
\stackrel{\displaystyle{K-\bar{K}}}{\Longrightarrow}\quad
\frac{\displaystyle{(\delta ^{LL} _{12})^2}}
{\displaystyle{ M^2_{\tilde{Q}} }}\le \frac{1}{(100 {\rm TeV})^2}\label{eq:KK} \eeq
where $\delta ^{LL} _{12}$ measures the departure from the identity matrix of  $m_Q ^2$  in
(\ref{eq:soft}) and   $M_Q $ is an average value for  $m_Q ^2$ \cite{GGMS96}. This shows the
{\it supersymmetric flavour problem}, i.e. the difficulties to solve
{\it simultaneously} the hierarchy problem ($m_{soft}$   $\le 1$  TeV) 
and NO FCNC's.

There are two 
 scenarios that can justify the lack of FCNC's in eq. (\ref{eq:KK}):
 {\it effective supersymmetry} \cite{Dimopoulos:1995mi} and 
 {\it Minimal Flavour violation} (MFV) \cite{MFV}. 
{\it Effective supersymmetry}
 still keeps naturalness by  allowing only the third
family of squarks to be below 1 TeV, 
then NP is expected in $\Delta B=2$-transitions (due to  $\delta ^{LL} _{13}$), 
while the first two families of squarks are decoupled, i.e. heavier than
5 TeV and  $\delta ^{LL} _{12}\sim 0$. This would be very exciting and a lot of
phenomenolgy could be accessible in the near future 
\cite{D'Ambrosio:2001zh,Fleischer:2003xx}.

However there is no obvious reason why the three families 
are so much different so that we have pursued also a
a different strategy to have New Physics 
at the TeV scale but no FCNC's: NP must obey some flavour symmetry  (MFV) so that
GIM mechanism it is still at work for the the three families. In supersymmetry, 
for instance, this global symmetry would strongly constrain the flavor matrices in 
(\ref{eq:soft}), so that (\ref{eq:KK}) turns in 
\cite{MFV,UUT0}:
\beq
{\cal L} _{\Delta F=2}=
 \frac{{C}}
{{\Lambda _{MFV}^2}}
[\frac{(V_{td}^* m _t ^2 V_{tb})^2}{v^4}
(\bar{d}_L \gamma ^\mu b_L)^2 + \frac{(V_{td}^* m _t ^2 V_{ts})^2}{v^4}
(\bar{d}_L \gamma ^\mu s_L)^2]\label{eq:MFVdF2}\eeq
Fixing ${C}\sim 1$ implies  a very strong bound on 
$\Lambda _{MFV}^2$. The flavour symmetry can be invoked in several contests. In fact
it was originally introduced in  Technicolour \cite{Georgi} to be protected from FCNC's: 
the underlying preonic dynamics, generating the vev's for the gauge and fermionic masses
should preserve the global flavour symmetry $G_{F}$ broken only by some  spurions 
\beq
G_{F}=\qquad  
\overbrace{{\rm U}(3)_Q \otimes {\rm U}(3)_U \otimes {\rm U}(3)_D \otimes
{\rm U}(3)_L \otimes {\rm U}(3)_E}^{\rm global\ symmetry} 
 +\qquad \overbrace{ Y_{U,D,E} }^{\rm spurions}
\label{eq:FS}\eeq
and spurion quantum numbers determined by  (\ref{eq:yuk}). This symmetry generates the 
${\cal L} _{ \Delta F=2}$ in  (\ref{eq:MFVdF2}). In order to solve the flavour 
problem this symmetry has been invoked
also in supersymmetry, gauge mediation \cite{gm}  and large extra dimensions
 \cite{rattazzi}
 ((see Fig. \ref{fig:3}).  
We have determined the general  dim-6 lagrangian consistent with the symmetry in 
(\ref{eq:FS}) in terms of some unknown coefficients $c_n$
\beq
 {\cal L} _{MFV}(Q,U,D,L,E,H)=
 {\cal L} _{\Delta F=2} 
+{ {\cal L}} _{\Delta F=1} =  \frac{1}{ \Lambda^2} \sum_{n} c_n {\cal O}_n ~+~{\rm h.c.} 
\eeq
Several intersting results have been obtained from this analysis: 
i) putting $ c_n \sim 1$ we can obtain 
strong constraints on $\Lambda$ from different processes, i.e. $\epsilon_K,\quad \Delta m_{B_d} 
\Longrightarrow \Lambda> 5{\rm TeV}$ \cite{MFV,UUT2} and 
$B\to X_s \gamma \Longrightarrow \Lambda> 8{\rm TeV}$
\cite{MFV}, ii) interesting new correlations among 
$B$ and $K$-physics, for instance 
 $\Br ( K_L\rightarrow \mu \bar{\mu}) \Longleftrightarrow  \Br ( B \rightarrow K l^+ l^-)$, etc.
to be tested in the very near future with the B- and K-factories
and  iii) if the CKM  matrix elements are known with  $5 \%$ accuracy, a measurement
at some percent level of 
$\Br ( K_L \to \pi^0  \nu\bar{\nu})$ has the chance to be the deepest probe of the SM;
in fact 
$\Lambda _{MFV}$  can be pushed  to $12{\rm TeV}$ \cite{MFV}. 
Still in  the MFV framework
but with two Higgses, i.e. in the case in which both Higgses are pretty light, 
 we have obtained all $\tan\beta$-enhanced Higgs-mediated FCNC's contributions.
Their effects in $B\to\ell^+\ell^-$,  $\Delta M_B$ and
$B\to X_s \gamma$ are particularly relevant \cite{MFV}.



\section{$K\rightarrow \pi \gamma \gamma$ decays and the
CP-conserving $K_{L} \rightarrow \pi ^0 \ell ^{+}\ell^{-}$ }
$K_{L} \rightarrow \pi ^0 \ell ^{+}\ell^{-}$ is a 
classic example of 
how our
control on low energy theory may help to disentangle 
short-distance physics.
In fact 
the effective current$\otimes $cur\-rent structure of weak interactions
obliges short-distance contributions to $K_{L}\rightarrow \pi ^{0}\ell
^{+}\ell ^{-},$ 
analogously to $K_{L}\rightarrow \pi ^{0}\nu \overline{\nu }$,
discussed in the previous section
 to be direct CP-violating \cite{reviews,chiralrev}. 
However, differently from the
neutrino case, $K_{L}\rightarrow \pi ^{0}\ell ^{+}\ell ^{-}$ 
receives also
non-negligible long-distance contributions:
 i) indirect CP-violating from 
one-photon exchange, discussed in the next section,
 and ii) CP-conserving from two-photon exchange, 
where the photons can be on-shell (two-photon
 discontinuity) and thus  directely related to the observable 
$K_{L}\rightarrow \pi ^{0}\gamma \gamma$ decay, or off-shell 
and then a form factor should be used \cite{DG95}. 
It is  possible  to avoid the potentially
large
background  from 
$K_{L}\rightarrow e^{+}e^{-} \gamma \gamma $ \cite{greenlee}
by studying time interferences  \cite{belyaev}.
\begin{figure}[t]
\begin{center}
\leavevmode
\epsfysize=3.5 cm
\epsfxsize=12 cm
\epsfbox{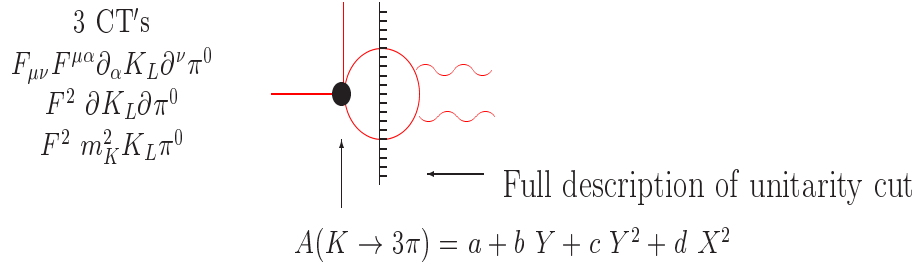}
\end{center}
\caption{$K_L\rightarrow \pi ^0 \gamma \gamma $: 
unitarity contributions from 
 $K \rightarrow 3 \pi$: $X,Y$ Dalitz var. \cite{CD93,CE93}}
\label{fig:klpggunit} 
\end{figure}
The present bounds (at 90\% CL)
from KTeV \cite{Littenberg:2002um,KTeVklpi0ll} are 
\begin{equation}
B(K_{L}\rightarrow \pi ^{0}e^{+}e^{-})<3.5\times 10^{-10}\quad
{\rm and}\quad
B(K_{L}\rightarrow \pi ^{0}\mu ^{+}\mu ^{-})<3.8\times 10^{-10}.
\end{equation}
\hspace*{0.1cm} The general amplitude for $K_{L}(p)\rightarrow \pi
^{0}\gamma (q_{1})\gamma (q_{2})$ can be written in terms of two 
Lorentz and gauge invariant amplitudes $A(z,y)$ and $B(z,y):$
\begin{eqnarray}
  && {\cal A}( K_{L}\rightarrow \pi ^{0}\gamma \gamma ) =  
   \frac{G_{8} \alpha }{ 4\pi }
   \epsilon_{1 \mu} \epsilon_{2 \nu} 
   \Big[  A(z,y) 
    (q_{2}^{\mu}q_{1}^{\nu }-q_{1}\ccdot q_{2}~ g^{\mu \nu } ) +  \nonumber \\
 && \qquad + 
   \frac{2B(z,y)}{m_{K}^{2}} 
    (p \ccdot q_{1}~ q_{2}^{\mu }p^{\nu} + p\ccdot q_{2} ~p^{\mu }q_{1}^{\nu}
   -p\ccdot q_{1}~ p\ccdot q_{2}~ g^{\mu \nu } 
   - q_{1}\ccdot q_{2}~ p^{\mu }p^{\nu } ) \Big]~, \quad
\label{eq:kpgg}
\end{eqnarray}
where $y=p (q_{1}-q_{2})/m_{K}^{2}$ and $z\,=%
\,(q_{1}+q_{2})^{2}/m_{K}^{2}$.
Then the double differential rate is given by 
\begin{equation}
\frac{\displaystyle \partial ^{2}\Gamma }{\displaystyle \partial y\,\partial %
z}\sim [%
\,z^{2}\,|\,A\,+\,B\,|^{2}\,+\,\left( y^{2}-\frac{\displaystyle \lambda
(1,r_{\pi }^{2},z)}{\displaystyle 4}\right) ^{2}\,|\,B\,|^{2}\,]~,
\label{eq:doudif}
\end{equation}
where $\lambda (a,b,c)$ is the usual kinematical function 
and $r_{\pi }=m_{%
\pi }/m_{K}$.
 Thus in the region of small $z$ (collinear photons) the $B$
amplitude is dominant and can be determined separately from the $A$
amplitude. This feature is crucial in order to disentangle the CP-conserving
contribution $K_{L}\rightarrow \pi ^{0}e^{+}e^{-}$. In fact  
the lepton pair 
produced by  photons  in $S$-wave, like 
an ${ A}(z)$-amplitude, are suppressed  by the lepton mass while
the photons  in $B(z,y)$ are
also in $D$-wave and so the resulting  $K_{L}\rightarrow \pi ^{0}e^{+}e^{-}$
 amplitude, 
$A(K_{L}\rightarrow \pi ^{0}e^{+}e^{-})_{CPC}$,  
 does not suffer from  the electron mass suppression \cite{kpee,EPR88}.

 The leading  $\mathcal{O}(p^{4})$ $K_L\rightarrow \pi ^0 \gamma \gamma$
amplitude \cite{Cappiello}
 is affected by two  large  $\mathcal{O}(p^{6})$ contributions: 
i) the full unitarity corrections from $K \rightarrow 3 \pi$
 \cite{CD93,CE93} in Fig. \ref{fig:klpggunit}  and   
ii) local contributions.
Fig. \ref{fig:klpggunit}  
enhances the $\mathcal{O}(p^{4})$ branching ratio
 by 
$40\%$ and generates a $B$-type amplitude. Local contributions
are generated by
three independent counterterms, as the one in Eq. (\ref{eq:Ksp6}),
 with the unknown coefficients 
$\alpha_1, \alpha_2$ 
and $\beta$ leading to contributions to $A$ and $B$  
in Eq. (\ref{eq:kpgg}) \cite{CE93}:
\begin{equation}
A_{\rm CT}=\alpha_1 (z-r_\pi^2)+\alpha_2, \quad
 B_{\rm CT}=\beta 
\label{eq:AB}.
\end{equation}


\noindent
\begin{figure}[t]
\vskip1cm
\begin{picture}(640,140)(0,0)
\put(0,0){\includegraphics[width=8cm,height=7cm]{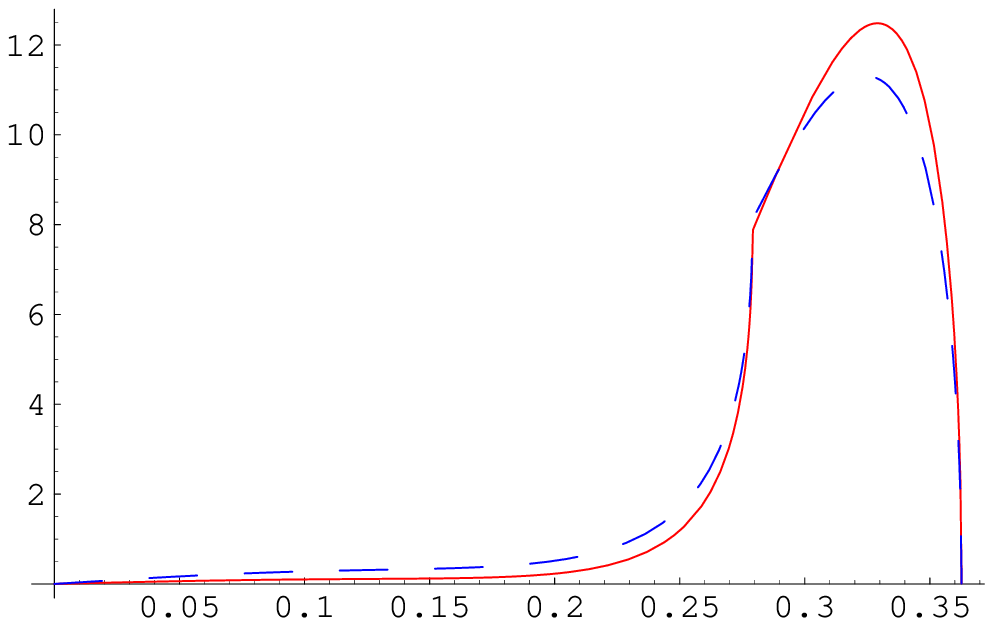}}
\put(30,100){\small B($K_L\rightarrow\pi^0\gamma\gamma $)}
\put(30,80){\small \red{$(1.36\pm 0.05)\cdot 10^{-6}$} }
\put(30,60){\small\blue{$(1.68\pm 0.1)\cdot 10^{-6}$}}
\put(80,180){ $ a_V$ } 
\put(80,165){\small \red{- 0.46$\pm$0.05 (NA48)}}
\put(80,150){\small  \blue{- 0.72$\pm$0.08 (KTeV)}}
\thicklines
\put(30,165){\red{\line(1,0){20}}}
\multiput(30,150)(20,0){2}{\blue{\line(1,0){10}}}
\put(250,0){\includegraphics[width=7cm,height=7cm]{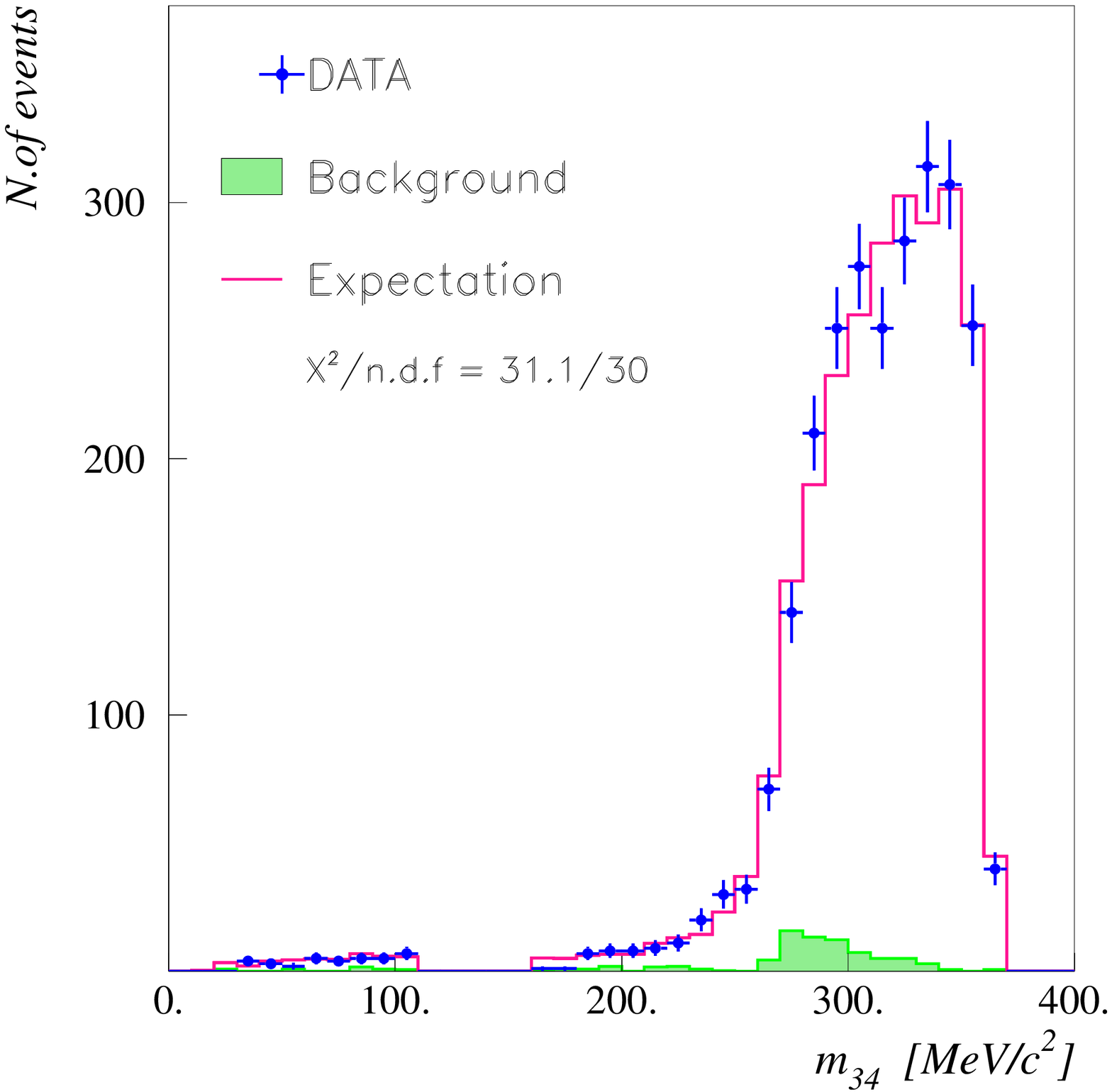}}
\put(300,110){\small \red{NA48}: No evts.}
\put(300,95){\small at low  $m_{\gamma\gamma}$}
\put(330,80){$\Downarrow$}
\put(290,60){B($K_L\rightarrow\pi^0 e^+ e^-$)}
\put(295,40){\red{$<5\cdot 10^{-13}$}}
\end{picture}
\caption{$K_L\rightarrow\pi^0 \gamma\gamma$ diphoton-invariant mass
spectrum for two values  of $a_V$:\red{ $-0.46$ (full curve)},  \blue{
$-$0.7  (dashed curve)}, corresponding respectively to the NA48 and KTeV measured value. 
On the right side the measured 
NA48 spectrum.}
\label{fig:klpggspectrum}
\end{figure}
If we assume 
  VMD \cite{SE88,EP90},
these couplings are related in terms of one constant, $a_V$:
\begin{equation}
\alpha_1=\frac{\beta}{2}=-\frac{\alpha_2}{3}=-4a_V.
\end{equation}
Though chiral counting suggests $\alpha_i ,\beta\sim 0.2 $,
VMD enhances this typical size. 
Actually a model, FMV, 
describing weak interactions of 
pseudoscalars ($\phi$'s)  
with vectors,
 ${\cal L}_W ^{FMV} (\phi,V^\mu)$,
based on factorization and couplings fixed by the Wilson coefficient
of the $Q_{-}$ operator,
 predicts:
\begin{equation} 
{\cal L}_W ^{FMV} (\phi,V^\mu) \quad \Longrightarrow 
\quad a_V=-0.6. \label{eq:FMV}
\end{equation}
Two experiments have measured these decays
in terms of one  parameter, $a_V$: KTeV \cite{KTeVkpgg99} and  NA48 
\cite{Lai:2002kf}. Their results and spectrum are shown in 
 from Fig. \ref{fig:klpggspectrum}
As we can see from Fig. \ref{fig:klpggspectrum} the spectrum at low $z$
is very sensitive to the value of $a_V$, or more generally to the size 
of the amplitude $B$ in Eq. (\ref{eq:kpgg}).
Since NA48 sees no evts. one can put  the bound shown in  Fig. 
\ref{fig:klpggspectrum} for this contribution.

\noindent
Recently Gabbiani and Valencia \cite{GV,Gabbiani:2002bk}  suggested to
fit the experimental 
$z$-spectrum (and the rate) with all three parameters in Eq. (\ref{eq:AB}). 
In fact,
VMD even in the best case is known to be only a good approximation and thus we 
think this  is a non-trivial
VMD test.

An important  issue is that while the size of 
$B(K_L\rightarrow \pi^0 e^+ e^-)_{CPC}
^{\gamma \gamma \ {\rm on-shell }}$
is an  issue that can be established firmly from the
$K_L\rightarrow\pi^0 \gamma\gamma$ spectrum, the contribution 
 when the two 
intermediate photons are off-shell is model dependent
and a form factor is needed \cite{DG95}. More theoretical work is needed and
probably a partial answer can come from the measurement of
$K_L\rightarrow\pi^0\gamma\gamma^*$ \cite{Gabbiani97,kpeeg}. 
Another interesting perspective  is the muon polarization in
$K_L\rightarrow \pi ^0  \mu ^+\mu^-$: here the $B-$type amplitude
is not dominant but the Greenlee background is more under control
\cite{diwan}.

\section{$K^\pm \rightarrow \pi ^{\pm} 
\ell ^{+}\ell ^{-}$ and $K_{S} \rightarrow
\pi ^{0}\ell ^{+}\ell ^{-}$ }
The CP-conserving decays $K^\pm (K_{S}) \rightarrow \pi ^{\pm} 
(\pi ^{0})\ell ^{+}\ell ^{-}$ are dominated by the 
long-distance process $K\to\pi\gamma \to \pi   \ell^+ \ell^-$ \cite{EPR}.
The decay amplitudes can in general be written in terms of one form 
factor $W_i(z)$ ($i=\pm,S$):
\begin{equation}
A\left( K_i \rightarrow \pi ^i \ell^+ \ell^- \right) = - \frac{%
\displaystyle e^2}{\displaystyle M_K^2 (4 \pi)^2} W_i(z) (k+p)^\mu \bar{u}%
_\ell(p_-) \gamma_\mu v_\ell(p_+)~,\label{eq:CPV_S}
\end{equation}
$z=q^2/ M_K^2$; $W_i(z)$  can be decomposed as the sum of 
a polynomial piece plus a
non-analytic term, $W_{i}^{\pi \pi}(z)$, generated  by  
the $\pi \pi $ loop, analogously  to the one in Fig. \ref{fig:klpggunit} 
for $K_L\rightarrow\pi^0\gamma\gamma$,  completely determined in terms 
of the physical $K\rightarrow 3 \pi$ amplitude \cite{DEIP}.
Keeping the polynomial terms up to  $\mathcal{O(}p^{6})$
we can write 
\begin{equation}
W_{i}(z)\,=\,G_{F}M_{K}^{2}\,(a_{i}\,+\,b_{i}z)\,+\,W_{i}^{\pi \pi
}(z)\;,
\label{eq:Wp6}
\end{equation}
where the  parameters $a_{i}$ and $b_{i}$ parametrize local 
contributions starting respectively at $\mathcal{O(}p^{4})$  
and $\mathcal{O(}p^{6})$. 
Recent data on $K^{+}\rightarrow \pi ^{+}e^{+}e^{-}$ and
$K^{+}\rightarrow \pi ^{+}\mu ^{+}\mu ^{-}$  by BNL-E865
\cite{kpllE865} have been successfully
fitted using 
Eq. (\ref{eq:Wp6})
and lead to
\begin{equation}
a_{+}\,=-0.587\pm 0.010,\qquad \,b_{+}=-0.655\pm 0.044~.  
\label{eq:ab+}
\end{equation}
Recentely HyperCP \cite{HyperCP} has attempted to measure   
the CP-violating width  charge asymmetry  in
 $K^{\pm}\rightarrow \pi ^{\pm} \mu ^{+}\mu ^{-}$ and 
it has found that it is consistent with 0
at 10\% level. Though the CKM prediction with accurate cuts
 is $\sim 10^{-4}$
 \cite{DEIP}, we are beginning to test new physics
affecting the operator $\bar{s} d\bar{\mu} \mu$ \cite{D'Ambrosio:2002fa}. 
The experimental size of the ratio $b_{+}/a_{+}$
exceeds the naive dimensional analysis estimate
$b_{+}/a_{+} \sim {\cal O}[M^2_K/(4\pi F_\pi)^2] \sim 0.2$, but can 
 be explained by a large VMD contribution.
Chiral symmetry alone does not allow  us to 
determine the unknown couplings $a_{S}$ and $b_S$ 
in terms of $a_{+}$ and $b_{+}$ \cite{EPR,DEIP}. Neglecting 
 the $\Delta I=3/2$ suppressed  
 non-analytic term $W_{S}^{\pi \pi}(z)$,  we obtain \cite{DEIP}
\begin{equation}
B(K_S \rightarrow \pi^0 e^+ e^-)  =  
\left[46.5 a_S^2 + 12.9 a_S b_S + 1.44 b_S^2 
\right]
\times 10^{-10} 
   \approx     5 \times 10^{-9} \times a_S^2~,
\label{eq:BRKS}
\end{equation}
The recent experimental 
information $B(K_S \rightarrow \pi^0 e^+ e^-) < 1.4 \times 10^{-7}$ 
\cite{NA48KS} let us derive the 
bound $|a_S| \le  5.3$; NA48 \cite{NA48KSW} 
and maybe KLOE \cite{KLOE} will assess in the near future the value 
of this branching at the least for values of $a_S$ of order 1. 
Of course even a strong bound is relevant, since it
 will ensure that this contribution is not dangerous  to measure direct CP
violation in
$K_{L}\rightarrow \pi ^{0}e^{+}e^{-}$. 
We remark that even a sizeable  $a_S$: 
 $a_S<-0.5$ or $a_S>1$, will lead to an interesting interference:
\begin{equation}
B(K_{L}\rightarrow \pi ^{0}e^{+}e^{-})_{CPV}\,=\,\left[
15.3\,a_{S}^{2}\,-\,6.8\frac{\displaystyle \Im \lambda _{t}}{\displaystyle %
10^{-4}}\,a_{S}\,+\,2.8\left( \frac{\displaystyle \Im \lambda _{t}}{%
\displaystyle 10^{-4}}\right) ^{2}\right] \times 10^{-12}~,
\label{eq:cpvtot}
\end{equation}
where $\lambda _{t}= V_{td} V_{ts}$.
The sign of the interference term is model-dependent, but, however is not 
a problem
to determine $\Im \lambda _{t}$ accurately   (up to a discrete
 ambiguity).

\section{Outlook}

Several other channels are interesting:
 $K_{L}\rightarrow \mu ^{+}\mu ^{-}$, for instance
is
 interesting 
to determine $V_{td}$. To this purpose an accurate knowledge
of the long distance contribution, $A_{LD}$, is required. 
Several theoretical methods  have been suggested to describe the relevant 
form factor, based on several theoretical arguments \cite{V98,DIP,Greynat:2003ja}. 
A virtue of the
the form factor in Ref. \cite{DIP} is to be testable. In fact the parameters and shape 
\cite{Littenberg:2002um,rad01} can be accurately measured and tested. 
Recentely   Ref.
\cite{Greynat:2003ja} has  pointed out that in the limit of large $N_c$, 
the form factor of 
Ref. \cite{DIP} may not completely satisfy QCD matching. 
However the corrections might be small.
Experimental results are promising \cite{Alavi-Harati:wd} and 
they have the chance to accurately test the appropriate form factor 
\cite{Littenberg:2002um,rad01}. 

\noindent
\begin{figure}[t]
\vskip1cm
\begin{picture}(640,140)(0,0)
\put(0,0){\includegraphics[width=8cm,height=7cm]{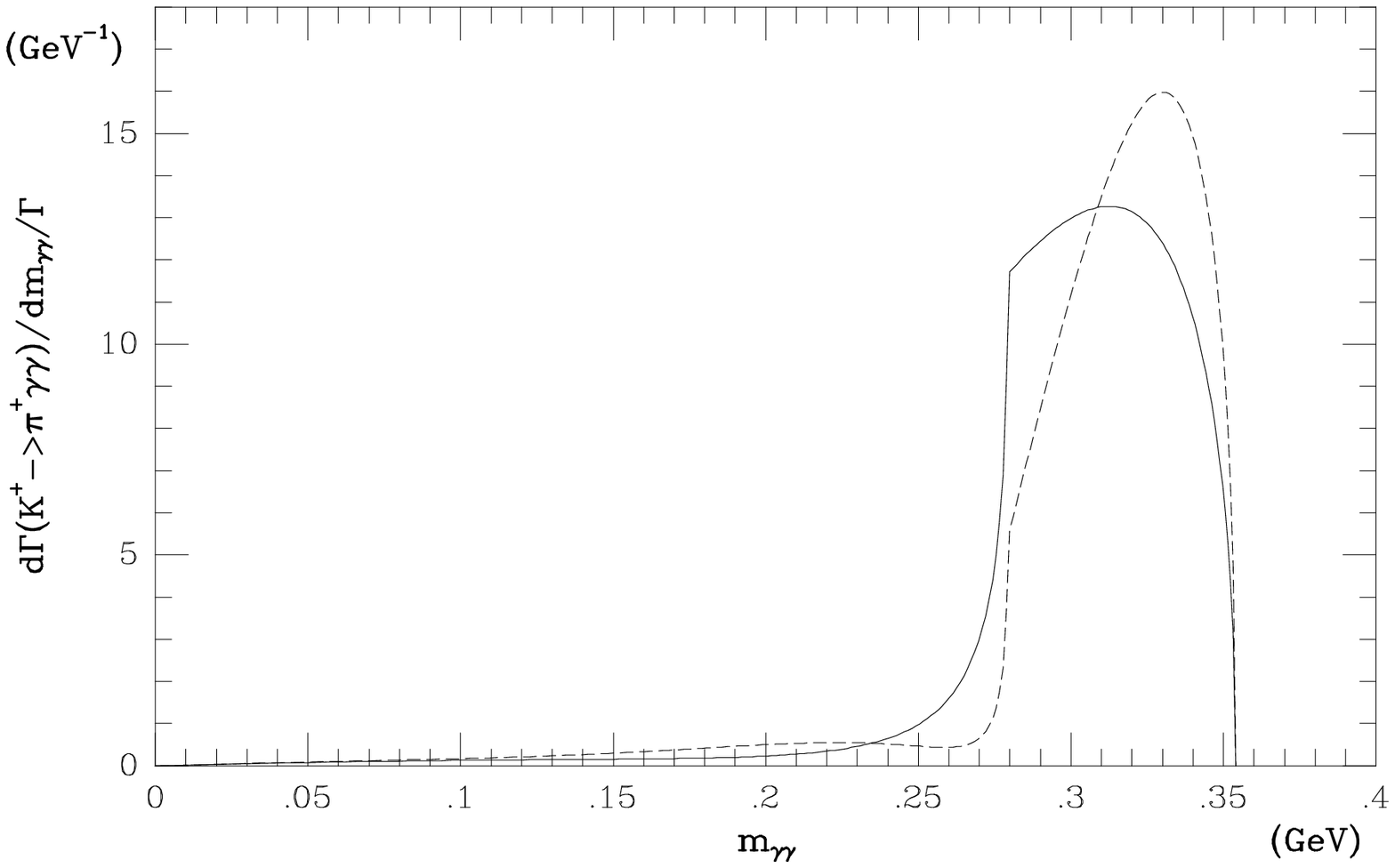}}
\put(30,180){\small B$(K^+\rightarrow\pi^+\gamma\gamma) \sim 1 \cdot 10^{-6}$ }
\put(80,140){ $ \hat{c}$ } 
\put(80,120){\small {0}}
\put(80,100){\small  {- 2.3 }}
\thicklines
\put(35,120){{\line(1,0){25}}}
\multiput(30,100)(20,0){2}{\line(1,0){10}}
\put(250,0){\includegraphics[width=7cm,height=7cm]{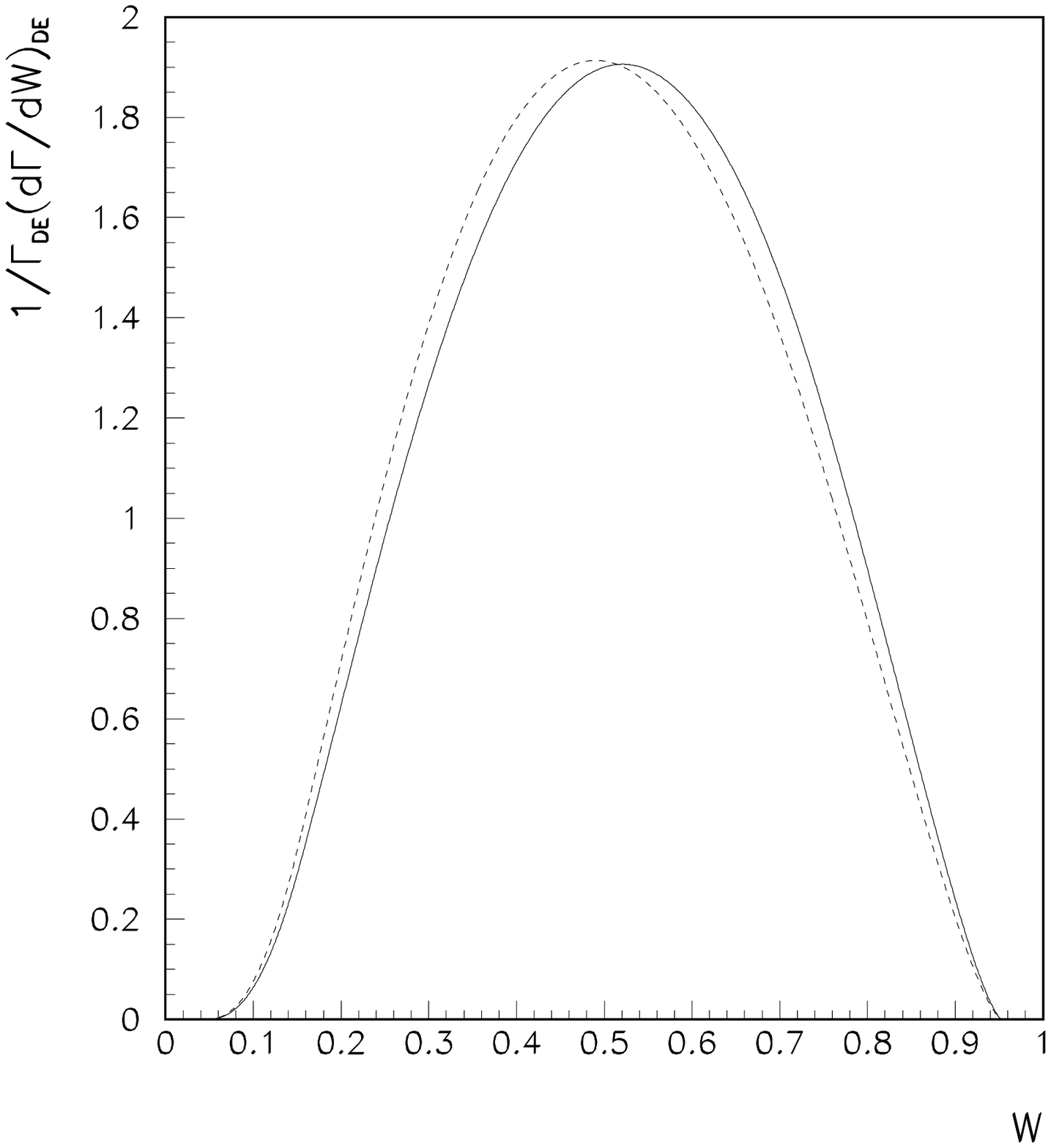}}
\put(315,50){\small $W^2=\frac{\displaystyle{(q\cdot p_K)(q\cdot p_+)}}
{\displaystyle{m_\pi^2 m_K^2}}$ }
\put(330,95){$K^+\rightarrow\pi^+ \pi^0\gamma$}
\end{picture}
\caption{On the left side
 $K^+\rightarrow\pi^+  \gamma \gamma$ diphoton-invariant mass
for two values of the  $\mathcal{O}(p^{4})$ parameter, $\hat{c}$, generated
by the lagrangian in  Eq.(\ref{eq:LW})
\cite{D'Ambrosio:1996sw}. On the right side 
$K^+\rightarrow\pi^+ \pi^0\gamma$
spectrum for two values  of the VMD parameters entering in $\mathcal{O}(p^{4})$
lagrangian in  Eq.(\ref{eq:LW}) \cite{DP98,gao00}. 
Regarding  the small difference among  the two curves in this plot, it is worthing 
pointing out  
that an analogous   difference  has been detected by KTeV in 
$K_L\rightarrow\pi^+ \pi^-\gamma$
\cite{KTeV-kppg} showing  large VMD also in the neutral channel}
\label{fig:kppgspectrum}
\end{figure}

NA48 has already
 seen $K_S \rightarrow \pi^0 \gamma \gamma $ \cite{Lai:2002dv}, however  larger
statistics is required in order to go beyond
the dominance of the $\pi ^0$-pole \cite{Cappiello}.
In the near future NA48  \cite{NA48KSW}
will give other interesting results on  rare $K_S$-decays: 
we are looking forward to their result on 
$K_S \rightarrow \pi ^0 e^+ e^-$.
Also  the improvements by NA48 \cite{NA48K+}
on the charge asymmetries should be exciting:
$K^+ \rightarrow \pi^+  \pi \pi$, $K^+ \rightarrow \pi^+ \pi^0 \gamma$
and $K^+ \rightarrow \pi^+ l^+ l^-$ \cite{DI98,D'Ambrosio:2002fa},
$K^+ \rightarrow \pi^+ \gamma \gamma$ \cite{Gao:2002ub};
this  will definetely lead also to improve
the CP conserving
decays  $K^+ \rightarrow \pi^+ \gamma \gamma$, 
$K^+ \rightarrow \pi^+ \pi^0  \gamma$ previously measured in Ref. \cite{BNL787}
and in  Ref. \cite{Aliev:2002tq} respectively: 
these channels have their own theoretical interest (see \cite{D'Ambrosio:1996sw} and 
\cite{gao00,ecker94})  and furnish additional
 information
to the neutral decays $K_L \rightarrow \pi ^0\gamma \gamma$ 
 and  
$K_L \rightarrow \pi^+ \pi^-  \gamma$. In particular we could have insights on the  
VMD in $\mathcal{O}(p^{4})$
lagrangian in  Eq.(\ref{eq:LW})\cite{EKW93,DP98}.

Interesting future experimental prospects have been exploited for instance in
Ref. \cite{Littenberg:2002um} and other may arise if for instance
NA48 will measure  in $K_{S}\rightarrow \pi ^{0}e^{+}e^{-}$ 
an interesting size for  
$a_S$ (see Eq. (\ref{eq:cpvtot})) \cite{belyaev}.

\section{Acknowledgements}
I wish to thank G. Buchalla, A. Ceccucci,
 G. Giudice, G. Isidori, C. Scrucca, A Strumia, M. Martini, I. Mikulec
for nice discussion and/or collaboration.
This work is supported in
part by TMR, EC--Contract No. ERBFMRX--CT980169 (EURODA$\Phi $NE).

\end{document}